\documentclass[12pt]{article}

\usepackage{epsfig}
\def \lsim{\mathrel{\vcenter
     {\hbox{$<$}\nointerlineskip\hbox{$\sim$}}}}

\newcommand{\beq}{\begin{equation}}
\newcommand{\eeq}{\end{equation}}
\newcommand{\beqa}{\begin{eqnarray}}
\newcommand{\eeqa}{\end{eqnarray}}
\newcommand{\beqar}{\begin{eqnarray*}}
\newcommand{\eeqar}{\end{eqnarray*}}

\newcommand{\labell}[1]{\label{#1}} 
\newcommand{\reef}[1]{(\ref{#1})}

\begin{document}

\thispagestyle{empty}

\hfill{}

\hfill{}

\hfill{CERN-TH/2001-259}

\hfill{hep-ph/0109287}

\vspace{32pt}

\begin{center} 
\textbf{\Large Cosmic Rays as Probes of Large Extra Dimensions and TeV
Gravity} 

\vspace{40pt}

Roberto Emparan$^{a,}$\footnote{Also at Departamento de F{\'\i}sica
Te\'orica, Universidad del Pa{\'\i}s Vasco, E-48080, Bilbao, Spain.},
Manuel Masip$^b$, and Riccardo Rattazzi$^{a,}$\footnote{On leave from
INFN, Pisa, Italy}

\vspace{12pt}

$^a$\textit{Theory Division, CERN}\\
\textit{CH-1211 Geneva 23, Switzerland}\\
\vspace{6pt}
$^b$\textit{Centro Andaluz de F\'\i sica de Part\'\i culas Elementales
(CAFPE)\\ and\\ Departamento de F{\'\i}sica Te\'orica y del
Cosmos}\\ \textit{Universidad de Granada, E-18071, Granada, Spain}\\
\vspace{6pt}
\texttt{roberto.emparan@cern.ch, masip@ugr.es, riccardo.rattazzi@cern.ch}
\end{center}

\vspace{40pt}

\begin{abstract} 

If there are large extra dimensions and the fundamental Planck scale is
at the TeV scale, then the question arises of whether
ultra-high energy cosmic rays might probe them. We study
the neutrino-nucleon cross section in these models. The elastic forward
scattering is analyzed in some detail, hoping to clarify earlier
discussions. We also estimate the black hole production rate. We study 
energy loss  from graviton mediated interactions 
and conclude that they can {\it not} explain the cosmic ray
events above the GZK energy limit. However, these interactions could
start horizontal air showers with characteristic profile and
at a rate higher than in the Standard Model.

\end{abstract}

\setcounter{footnote}{0}

\newpage

\section{Introduction}

In this paper we explore the possibility that the primary particles for
ultrahigh-energy cosmic rays are neutrini interacting gravitationally
with atmospheric nucleons. An obvious objection to this idea is that
the gravitational interaction is too weak to produce any sizable cross
section for this process. However, this point needs to be fully
reconsidered in theories where the fundamental scale is around the TeV,
as postulated in models involving large extra dimensions and a
four-dimensional brane-world \cite{ADD,RS1}. In these scenarios not
only does the cross section for elastic gravitational scattering
increase at cosmic ray energies, but there is also the possibility that
the collision results into the formation of microscopic black holes.
Both effects can dramatically increase the cross section for scattering
between neutrini and atmospheric nucleons, and hence they may play a
role in explaining the most energetic cosmic ray events. 

In the recent past, this possibility has been entertained in a number
of papers, with differing conclusions \cite{add2,NuSh,cosmic}. There
has been a controversy as to the right way to perform the calculations,
and how to implement unitarity at high energies. We hope to shed some
light on these issues, and show that, actually, the situation is quite
simple once the appropriate point of view is taken \cite{bwwaves}.
Besides, the possibility of producing black holes in cosmic ray
collisions needs to be addressed in detail. Once we have, hopefully,
settled the terms for the analysis, we will turn to the actual
discussion of whether the first signatures from low-scale unification
and large extra dimensions might come from the study of
ultrahigh-energy cosmic rays (UHECR).

While this work was in progress, detailed analyses of the possibility of
black holes forming at the LHC have appeared \cite{DL,GTh,DE}. During
the final stage of this work, a paper \cite{FS} has appeared which
studies the production of black holes in cosmic rays, and also
investigates their detection in horizontal air showers. Our work
complements that of ref.~\cite{FS}: the latter focuses on the
phenomenology of the detection of black holes, whereas we address in
more detail the theoretical aspects of neutrino-nucleon scattering in
TeV-gravity theories. A paper discussing further aspects of the
detection of these showers has appeared when this work was ready for
submission \cite{GA}.

\section{Ultra-high energy scattering on the brane}

An essential feature of the gravitational interaction is that at
center-of-mass (c.o.m.) energies $\sqrt{s}$ well above the fundamental 
scale, the coupling to gravitational coupling grows so large that
graviton exchange dominates over all other interactions. This is
actually the case for atmospheric nucleons being hit by neutrini of
energy  $E_\nu \sim 10^{11}$ GeV ($\sqrt{s}\sim 10^6$ GeV) if the
fundamental scale is around 1 TeV. In particular, if the impact
parameter $b$ is sufficiently smaller than the radius of compactification, 
the extra-dimensions can be treated as non-compact. In this regime
one would be probing the extra dimensions purely by
means of the gravitational interactions. 

Another consequence of ultra-high energies in gravitational scattering
is that, to leading order, its description involves only classical
gravitational dynamics. In particular, this means that we do not need
any detailed knowledge of quantum gravity to perform the calculations:
any theory that has General Relativity as its classical limit should
yield the same results\footnote{This has been noted often earlier,
e.g., in \cite{tH,MuSoACV,BaFi,bwwaves,DL,GTh}.}. One can distinguish
different regimes in the scattering (and we will do so below), but
perhaps the most spectacular effect at such energies is the expected
formation of black holes, via classical collapse, when the impact
parameter is of the order of the horizon radius of the (higher
dimensional) black hole \cite{MyPe}\footnote{We define $M_D$ as in
\cite{GRW}.},
\beq
\label{bhradius}
R_S=\left({2^n\pi^{n-3\over 2}\Gamma\left({n+3\over 2}\right)
\over n+2}\right)^{1\over n+1}
\left({s\over M_D^{2n+4)}}\right)^{1\over 2(n+1)}\,.
\eeq
This implies that any dynamics at $b<R_S$ is completely shrouded by the
appearance of trapped surfaces: Ultrashort distances are directly probed
only for energies around the fundamental energy scale. 

In the following we will assume for simplicity that no scale for new
physics arises before reaching the scale for the fundamental energy
$M_D$. In particular, we assume that scales such as the string tension,
or the tension and thickness of the brane, do not appear before that
scale. This prevents the possibility of additional effects arising at
impact parameters larger than the ones that give rise to black hole
formation. If this is the case, then the picture for ultrahigh-energy
scattering that we describe here should be largely universal.
Nevertheless, stringy effects below the regime where General Relativity
can be trusted may be readily accommodated \cite{DE} and should not
introduce large changes in our results. 

Ultra-high energy scattering in the Randall-Sundrum model has been
addressed in \cite{bwwaves}, and the different regimes for the
scattering in the present case are qualitatively the same as described
there. At large impact parameters one does not expect formation of
black holes, but in this case, the leading contribution to the
scattering amplitude is exactly (non-perturbatively) calculable within
an eikonal approach \cite{tH,MuSoACV,KaOr}. This is known to work
particularly well for high energy gravitational scattering at large
impact parameter \cite{MuSoACV,kabat}\footnote{Loops involving only momenta of
internal gravitons are suppressed by factors of $1/(M_Db)^2$.}. 

The eikonal resummation of ladder and crossed-ladder diagrams is
achieved by computing the scattering amplitude as
\beqa
\mathcal{M}(s,t)&=&{2 s\over i}\int d^2b\; e^{i\mathbf{q}\cdot\mathbf{b}}\;
\left(e^{i\chi (s,b)}-1\right) \nonumber \\
&=&{4\pi s\over i}\int db\;bJ_0(b q)\;\left(e^{i\chi (s,b)}-1\right)\,.
\labell{eikonampl}
\eeqa
This amplitude is well defined for any values of the exchanged 
momentum $q=\sqrt{-t}$ ($t<0$ since the scattering is elastic).  The
eikonal phase  $\chi(s,b)$ is obtained from the Fourier-transform to
impact parameter space of the Born amplitude. Alternatively, it can be
obtained from the deflection of a particle at rest when crossing the
gravitational shockwave created by a second particle \cite{tH}.  

Note that the transforms in impact parameter space are two-dimensional,
since the particles scatter in three spatial dimensions. Nevertheless,
the exchanged gravitons propagate in the $4+n$ dimensional space.
Moreover, we are working at a scale where the spectrum of Kaluza-Klein
modes is essentially continuous. 
In this
case the Born amplitude comes out easily as \cite{GRW}
\beq
i\mathcal{M}_{Born}=i\pi^{n/2}\Gamma\left(1-{n\over 2}\right)
{s^2\over M_D^{2+n}} (-t-
i\epsilon)^{{n\over 2}-1}\,.
\label{bornm}
\eeq
Hence the eikonal phase\footnote{This corresponds to the linearized
approximation to \reef{eikonampl}.},
\beq
\chi(s,b)={1\over 2is}\int {d^2
q\over(2\pi)^2} e^{i\mathbf{q}\cdot\mathbf{b}}i\mathcal{M}_{Born}\,,
\eeq
which is finite for $b\neq 0$, is
$\chi(s,b)=(b_c/ b)^n$,
where we have defined
\beq
b_c^n=
{(4\pi)^{{n\over 2}-1}\over 2}\Gamma\left({n\over 2}\right)
{s\over M_D^{2+n}}\,.
\label{bc}
\eeq

Having the phase $\chi(s,b)$ is sufficient for numerical
evaluation of the eikonalized amplitude \reef{eikonampl}. The result in
eq. \ref{eikonampl} can be written in terms of Meijer functions.
However, it is easy to get simple analytical expressions for the amplitude 
in both regimes of $qb_c\gg 1$ and $qb_c\ll 1$..
When $q\gg b_c^{-1}$ the phase $\chi(s,b)$ yields a sharp
peak for the eikonal amplitude in \reef{eikonampl}, which allows for an
evaluation near the saddle point $b_s = b_c(qb_c/n)^{-1/(n+1)}\ll
b_c$:
\beqa
\mathcal{M}_{saddle}&=&{4\pi i e^{i\phi}\over\sqrt{n+1}}\left[
(4\pi)^{{n\over 2}-1}\Gamma\left({n\over 2}+1\right)\left({s\over
qM_D}\right)^{n+2}\right]^{1\over n+1}\nonumber\\
&\equiv&Z_n\left({s\over
qM_D}\right)^{n+2\over n+1}\,.
\labell{eikonsaddle}
\eeqa
The phase $\phi=(n+1)(b_c/b_s)^n$ 
is real for $t<0$. Observe that the amplitude is non-perturbative in the
gravitational coupling $1/M_D^{2+n}$.
In the limit $q\to 0$ one gets instead
\beq
{\cal M}(q=0)=2\pi i\; s b_c^2\; \Gamma\left(1-{2\over n}\right)e^{-i\pi\over n}\,,
\eeq
which is finite for $n>2$. For $n=2$ the real part of ${\cal M}$ has a
logarithmic singularity
\beq
{\cal M}\stackrel{q\to 0}{ =} -4\pi s b_c^2\; \ln (qb_c).
\label{logsing}
\eeq
Notice that also at small $q$ the amplitude is non-analytic in the gravitational
coupling. Indeed the amplitude at $q\to 0$ is effectively described
by the (Born) operator ${\cal T}$ defined in ref. \cite{GRW} but with an effective UV 
cut-off
$\sim b_c^{-1}$ on the mass of the exchanged KK modes. This cut-off originates
from the interference with the multigraviton exchange diagrams in the eikonal
series.

The eikonal amplitude will be used in the next section to compute the
differential cross section for neutrino-nucleon scattering.
At the partonic level we have
\beq
{d\sigma\over dq^2}={1\over 16\pi s^2}|\mathcal{M}|^2\,.
\label{diffcross}
\eeq
We can also derive 
the total elastic cross section from the  optical theorem:
\beq
\sigma_{el}={\mathrm{Im}\mathcal{M}(q=0)\over s}
=2\pi b_c^2\;\Gamma\left(1-{2\over n}\right)\cos{\pi\over n}\,,
\eeq
i.e., it is essentially given by the area of a disk of radius 
$\sim b_c$. 

Observe that
\beq
\sigma_{el}\sim s^{2/n}\,.
\eeq
This growth of the cross section at high energy is slower than the 
perturbative result $\sigma \sim s^2$, and also slower (for $n>2$) than
the linear dependence $\sigma \sim s$ postulated (apparently for all
$n$) in \cite{NuSh}. Unitarity in impact parameter space is manifest in
the eikonal amplitude \reef{eikonampl}\footnote{Notice that in order to
achieve unitarity we have needed to perform an all-order loop
resummation. As argued in \cite{MuSoACV}, this is essential when
considering energies above $M_D$. This point is missed in some of the
earlier work, such as the last reference in \cite{cosmic}.  Note as
well that the Froissart bound \cite{Fr}, generalized to higher
dimensions in \cite{ChF}, does not apply since the exchanged particle
is massless.}. For large impact parameter this implies as well
unitarity for high partial waves. Partial wave unitarity at shorter
impact parameter is a harder problem, and indeed, corrections to the
eikonal amplitude are expected to become crucial. As $t$ grows,
graviton self-interactions, which carry factors of $t$ associated to
the vertices, increase the attraction among the scattered particles,
and it is expected that, eventually, gravitational collapse to a black
hole will take place. Hence the initial state is expected to be
completely absorbed, but in such a way that any short distance effects
will be screened by the appearance of a horizon. Indeed as shown in
ref.\cite{MuSoACV} the effects of the non-linearity of gravity are
suppressed by a power of $R_S/b$, so our eikonal approximation should
be valid for $b\gg R_S$ and its breakdown be associated to the
formation of black-holes. This relation between eikonal breakdown and
black-hole formation can also be establisehd as follows. In the region
$b\ll b_c$, there is a one to one correspondence
between the transferred momentum $q$ and the saddle point impact
parameter $b_s$. The case $q\sim \sqrt s$, where the (small angle)
eikonal approximation breaks down, corresponds precisely to $b_s\sim
R_S$. Notice in passing that we can also write eq. (\ref{diffcross}) as
$d\sigma=2 \pi b_s\,db_s$, as expected for a classical trajectory with
impact parameter $b_s$.

At present, the cross section for black hole production can only be
estimated as the geometric cross section,
\beq
\sigma_{bh}\sim\pi R_S^2\,.
\eeq 
with $R_S$ as in \reef{bhradius}. In this case $\sigma_{bh}\sim
s^{1/(n+1)}$, again slower than linear.

Clearly this result cannot be very accurate. Radiation is expected to
be emitted during the collapse, and the amount of energy that is
expected to be radiated in the process can be a sizable fraction of the
total energy (perhaps around $15-30\%$, from four-dimensional estimates
\cite{DEa}), but at large enough energies it will not be able to
prevent the collapse. This effect will tend to reduce the above value
for the cross section. However, there are also factors which increase
it, such as the fact that a black hole acts as a somewhat larger
scatterer ($40-75\%$ larger radius \cite{EHM3}). It seems reasonable to
expect that the above expression is not off by any large factors.
Finally, note that these black holes form through classical collapse.
In \cite{voloshin} the semiclassical instanton contribution to the
nucleation of black holes was considered. Being a tunneling process, it
is exponentially suppressed. Hence, it can be neglected relative to the
(real time) classical collapse we are considering.

\section{Neutrino-nucleon scattering and black hole
production}

These results can now be readily applied to neutrino-nucleon scattering
at ultra-high energies. At impact parameters $b<1$ GeV$^{-1}$ the
neutrino interacts essentially with the partons, and if $b>R_S$ the
eikonal approximation gives a good description of the scattering. At
smaller distances, trapped surfaces are expected to form and the
neutrino and the parton will collapse to form a black hole. 

In order to numerically evaluate the amplitude \reef{eikonampl}, we
proceed as follows. First, we write it as
\begin{equation}
i{\cal M}=4\pi s b_c^2 \int x dx J(x qb_c)(e^{i/x^n}-1)=4\pi s b_c^2 
\hat{\cal M}(qb_c)\,.
\end{equation}
At large values of $qb_c$ we know this is well described by the simple
result \reef{eikonsaddle}. It is convenient to extract
this behavior, and write the
squared amplitude $|\hat{\cal M}|^2$ as
\begin{equation}
|\hat{\cal M}|^2= \left (1+(qb_c)^2\right )^{-\frac{n+2}{n+1}}\,
\frac{n^{\frac{2}{n+1}}}{n+1} \,F(qb_c)\,.
\end{equation}
The prefactors have been chosen in such a way that for
$qb_c\to \infty$ the function $F$ goes to 1. Apart for the case $n=2$ where
it has a mild logarithmic singularity at $qb_c\to 0$ (see eq.~(\ref{logsing})), 
$F$ is $O(1)$  over the full range of $qb_c$.

For our applications it is useful to study the
cross section as a funtion of the fraction
$y$ of energy transferred to the nucleon:
\beq
y={E_\nu - E_\nu '\over E_\nu} = {q^2\over x s}\;.
\eeq
where $x$ is the fraction of proton momentum carried by the parton.
Summing over partons we have 
\beq
\frac{d\sigma}{dy} = \int_0^1 dx\; {1\over 16\pi x s}\;
\left(\sum_i f_i(x,\mu) \right)\; |{\cal M}(x,y,{\sqrt s}/M_D)|^2.
\eeq
Here $f_i(x,\mu)$ are the parton distribution functions (PDFs) (we use
the CTEQ5 set extended to $x < 10^{-5}$ with the methods in
\cite{Gandhi:1996tf}). Notice that quarks and gluons interact in the
same way. The scale $\mu$ should be chosen in order to minimize the
higher order QCD corrections to our process. A simple, but naive, choice
would be $\mu= q$. However $1/q$ does not really represent the typical
time or length scale of the interaction. As we have seen, in the
stationary phase regime, the neutrino is truly probing a distance
$b_s\gg 1/q$ from the parton. Heuristically: the total exchanged
momentum can be large, but through the exchange of many soft gravitons.
So we believe that a better normalization is to take $\mu=b_s^{-1}$ when
$q>b_c^{-1}$ and $\mu=q$ if $q<b_c^{-1}$. The latter choice is
effectively equivalent to choosing $\mu=b_c^{-1}$ as at small $q$ the
eikonal corresponds to a pointlike interaction. Our choice of $\mu$ is
consistent with the fact that gravity at ultra-Planckian energies is
dominated by long distance classical physics. Choosing $\mu=q$ would
also make little sense. $q$ can be as big as $\sim \sqrt s\gg M_D$, but
the evolution of the PDF's at $Q^2>M_D^2$ cannot be simply performed
withing QCD, as truly quantum gravitational effects (string theory)
would come into play. Instead as $\sqrt s$ grows above $M_D$, and $t/s$
is kept fixed but small, the impact parameter $b_s$ grows and we are
less sensitive to short distance physics. As a matter of fact, for large
enough $s$ the total $\sigma_{\nu N}$ will be bigger than the proton
area $\sim ({\rm GeV})^2$: at higher energies the parton picture breaks
down, the proton interacts gravitationally as a pointlike particle, and
the neutrino scatters elastically on it.

A useful quantity to study is the   cross section integrated for $y>y_0$.
In Fig.~1 and Fig.~2 we plot this quantity
 for $M_D=1$ TeV and $M_D=5$ TeV, respectively.
We include the cases with $n=(2,3,6)$ and
$E_\nu=(10^{10},10^{12},10^{14})$ GeV.

\begin{figure}[ht]
\begin{center}\leavevmode  %
\epsfbox{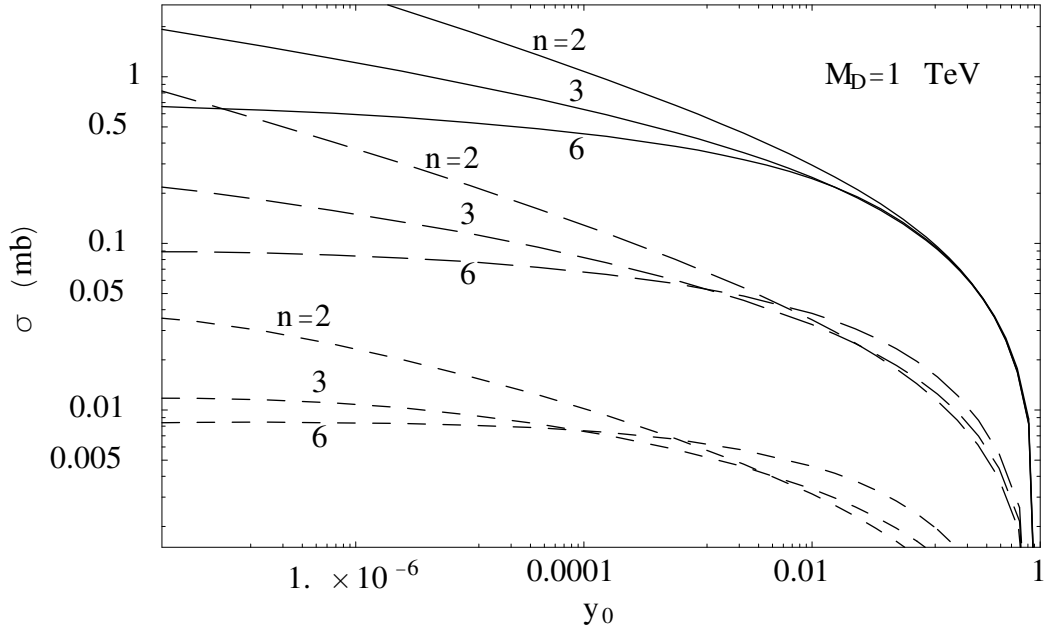}
\caption{ Elastic cross section vs.\ minimum fraction of energy  lost
by the neutrino for $M_D=1$ TeV and $n=2,3,6$ large extra dimensions.
Solid, long-dashed and short-dashed lines correspond respectively to
$E_{\nu}=10^{14}$ GeV, $E_{\nu}=10^{12}$ GeV and $E_{\nu}=10^{10}$
GeV. 
\label{Fig. l}}
\end{center}
\end{figure}

\begin{figure}[ht]
\begin{center}\leavevmode  %
\epsfbox{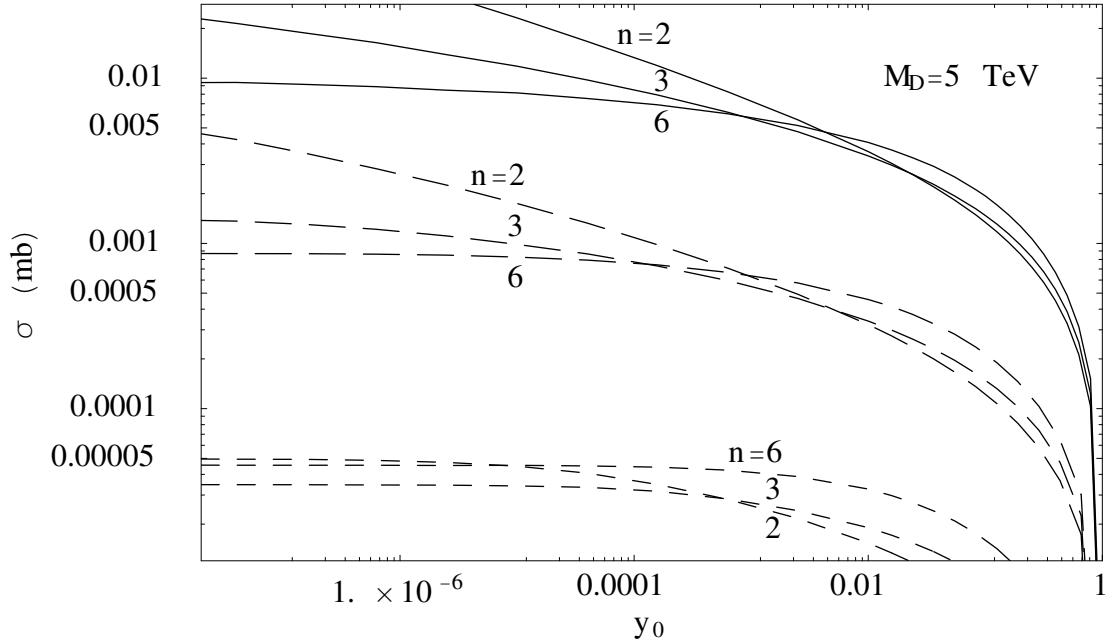}
\caption{As in Fig.~1 but for $M_D=5$ TeV.
\label{Fig. 2}}
\end{center}
\end{figure}

Finally, to estimate the total cross section to 
produce a black hole in a neutrino-nucleon 
scattering we compute
\beq
\sigma=\int_{M_D^2/s}^{1} dx\;
\left(\sum_i f_i(x,\mu) \right)\;
\pi R_S^2,
\eeq
where $R_S$ is given in Eq.~(\ref{bhradius}) and $\mu=R_S^{-1}$. Again
for the choice of scale in the PDF's the previous discussion applies:
the Schwarzschild radius rather than the black-hole mass sets the time
scale of gravitational collapse. Notice that in a more standard case of,
say, neutrino-quark fusion into an elementary lepto-quark the right
choice would be $\mu$ of the order of the lepto-quark mass. The crucial
difference is that the black-hole is not an elementary object: its
physical size is much bigger than its Compton wavelength.

We plot in Fig.~3 this cross section versus the energy of the incoming
neutrino for $n=(2,3,6)$ and $M=(1,5)$ TeV. We include plots with
$xs>M_D^2$ (solid) and $xs>(10 M_D)^2$ (dots). These correspond to the
cross sections for producing black holes with a mass larger than $M_D$
or $10M_D$, respectively. The sizeable difference between the two choices
of a minimum $x$, indicates that the production of light (small) black holes dominates:
the fast decrease with $x$ of the p.d.f.'s wins over the growth $\propto x^{1/(n+1)}$
of the partonic BH cross section. Notice, on the other hand, that the total elastic cross
section is less dependent on the small $x$ region  and is 
dominated by $x\sim 1$ for the case $n\leq 3$. This is because of the faster growth
$\propto x^{2/n}$ of the partonic elastic cross section.

\begin{figure}[ht]
\begin{center}\leavevmode  %
\epsfbox{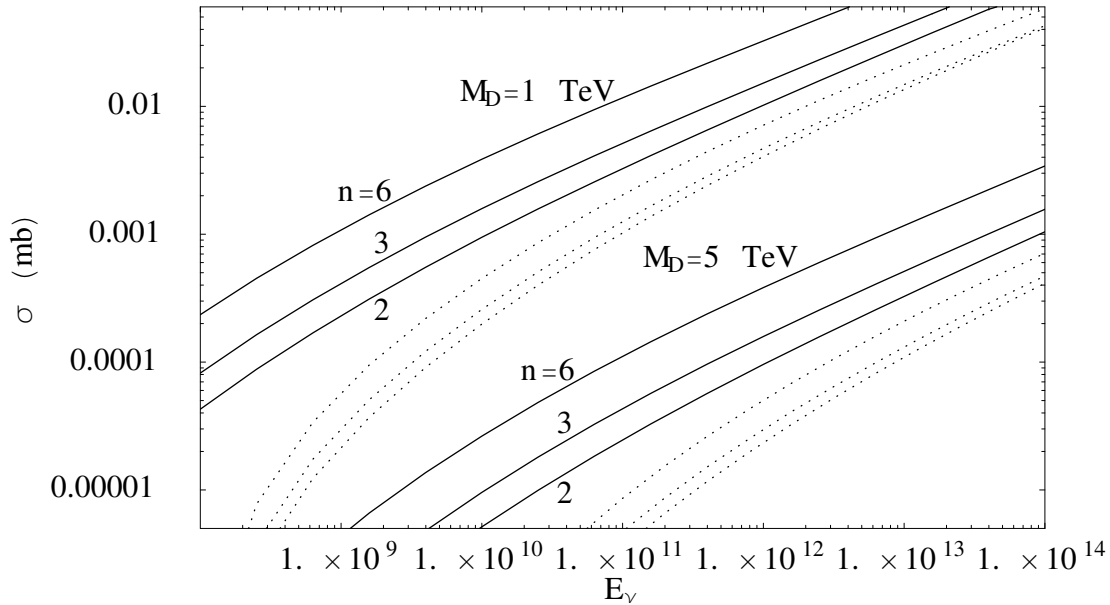}
\caption{
Cross section for black hole production as a function of $E_\nu$, for
$M_D=1,5$ TeV and $n=2,3,6$. Solid and dotted lines correspond to
$xs>M_D^2$  and $xs>(10 M_D)^2$ respectively.
\label{Fig. 3}}
\end{center}
\end{figure}

\section{Discussion}

We are now ready to discuss the implications of our results on the
phenomenology of ultra-high energy cosmic rays. The first question is
whether neutrino nucleon scattering at super-Planckian energies can
explain the observed cosmic ray events with energy $E>E_{GZK}=5 \times
10^{10}$ GeV. It is known since long ago that cosmic protons with
energy above $E_{GZK}$ are damped by inelastic scattering with the
microwave background photons. The relevant reaction is $p+\gamma \to
p+\pi$, and $E_{GZK}$ is the threshold proton energy given the photon
temperature. Because of this reaction,  ultra energetic cosmic protons 
are brought down to $E\simeq E_{GZK}$ within a few  Mpc. Since there
are good reasons to believe that the cosmic protons have extra-galactic
origin, we should observe a sharp drop in the observed event rate at
$E>E_{GZK}$. However, various experiments do not observe this drop at
all. There have been several suggestions to explain that. One idea is
that the primary particles for the UHECR are neutrini
\cite{neutrini,Tyler:2001gt,NuSh,cosmic}, as these particles interact
negligibly with the microwave background and are essentially undamped.
However, any of these suggestions has to face the fact that, within the
Standard Model (SM), the neutrini interact too weakly also with the nucleons in
the atmosphere. In order to explain the ultra-GZK events by cosmic
neutrini one needs new physics enhancing their cross section with
nucleons at high energy. In ref.~\cite{NuSh} it was suggested that, in
models with TeV scale gravity, the eikonalized cross section could be
of the right order of magnitude. However ref.~\cite{NuSh} did not
investigate the rate of energy loss in the eikonalized process, and, in
particular, did not pay attention to its ``softness''. The production 
of black holes was also neglected in ref.~\cite{NuSh}. 

As a matter of fact, in order to determine the signal it is important
to establish quantitatively which is the process that dominates energy
loss -- whether elastic gravitational scattering or black hole
production. It turns out that energy loss is mostly determined by black
hole production and by scattering at $y\sim 1$. (As we already pointed
out, and as can be seen from comparing Figs.~1,2 with Fig.~3, the
gravitational cross section at $y\sim 1$ becomes comparable to
$\sigma_{bh}$, though its precise value is not calculable within our
linearized gravity approximation.) To see this, consider a neutrino
travelling through a medium  of density $\rho$. The mean free path for
black hole production, at which all energy is lost  to the
shower, is $L_{bh}=(\sigma_{bh} \rho)^{-1}$. While travelling through
the  medium the neutrino also loses energy through the softer, but more
frequent,
eikonalized scattering. After travelling a distance $L_{bh}$, the
energy fraction lost to soft scatterings with $y<y_0$ is controlled by
the quantity
\begin{equation}
\label{etadef}
\eta(y_0)=\int_0^{y_0}y\frac{d\sigma}{d y}\,\rho\,
L_{bh}\;dy=\frac{1}{\sigma_{bh}}
\int_0^{y_0}  y\frac{d\sigma}{d y}\;dy.
\end{equation}
When $\eta$ is less than 1 the soft scatterings play
a negligible role in the transfer of energy to the atmosphere.
In Fig.~4 we plot $\eta$ for several cases: they all show that black
hole formation and scattering at large $y$ dominate energy loss.
Notice that, by the discussion at the end of the previous section, energy loss
is thus dominated by parton scatterings with
${\sqrt{x s}}\sim M_D$, {\it i.e.} in the Planckian regime.

\begin{figure}[th]
\begin{center}\leavevmode  %
\epsfbox{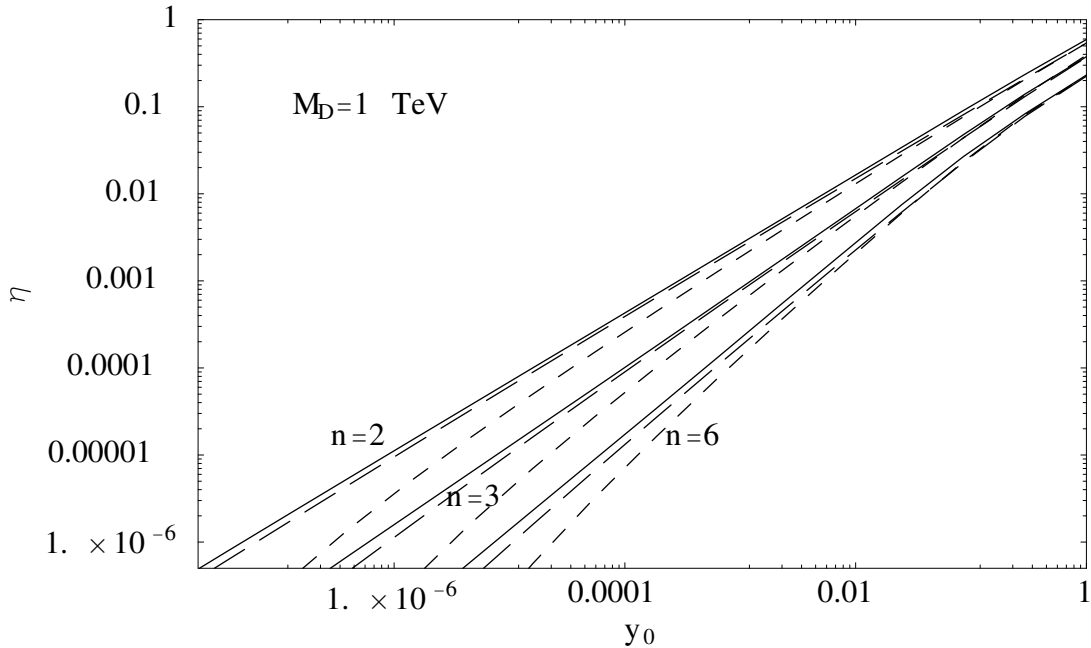}
\end{center}
\caption{Fraction $\eta$, defined in eq.~\reef{etadef}, of neutrino
energy lost to soft scatterings. Solid, long-dashed, and short-dashed
lines correspond to $E_\nu=10^{14}$, $10^{12}$ and $10^{10}$ GeV,
respectively.
}
\label{Fig.4}
\end{figure}

The observed showers above the GZK cut-off are all consistent with an
incoming particle that loses all its energy to the shower already in
the high atmosphere. From the above discussion, low scale gravity 
could explain these events if the mean free path $L_{bh}$ for black
hole production were somewhat smaller than the vertical depth of the
atmosphere. In standard units, the vertical depth $x_v$ is measured as
the number of nucleons  per unit area $x_v=1033 \times
N_A/\mathrm{cm}^{-2}=\mathrm{mb}^{-1}$ (where $N_A$ is the Avogadro
number), so the requirement is $\sigma_{bh}>{x_v}^{-1}=$mb. From Fig.\ 3
one can see that, at the relevant energies, the black hole cross
section, however large, falls short of this requirement. In order to
satisfy $\sigma_{bh}>$mb the gravity scale $M_D$ should be well below a
TeV, which would contradict collider limits.

Hence, we conclude that neutrino-nucleon interactions in TeV-gravity
models are \textit{not} sufficient to explain the showers above the GZK
limit. Also, at present, cosmic rays do not appear to place any
significant bounds on such scenarios.

Nevertheless, neutrino-nucleon cross sections $\sigma_{\nu N}$ in the
range $10^{-5}$ mb to 1 mb, like in our scenario, can still lead to
interesting new phenomena in cosmic ray physics, which may be observed
in upcoming experiments. Cosmic primaries with cross section below 1 mb
can travel deep into the atmosphere before starting a shower. In
particular they can cross the atmosphere at a large zenith angle and
start characteristic horizontal air showers\footnote{For the
observation of horizontal air-showers from neutrino primaries, see
\cite{HAS}.}. The horizontal depth of the atmosphere is $x_h$ is about
36 times the vertical one, so that for $\sigma_{\nu N}\lsim .1$ mb a
neutrino can travel horizontally down to the interaction point.  In the
Standard Model the charged-current cross section is $\sigma_{\nu N}\sim
10^{-5} (E/10^{10}{\rm GeV})^{0.363} $ mb. No horizontal air shower has
been detected so far. However, conservative estimates of the flux of
ultra energetic cosmic neutrini \cite{Tyler:2001gt} suggest that the
next generation of experiments should be barely sensitive to neutrino
cross sections of the order of the SM one. In our scenario $\sigma_{\nu
N}$ can be considerably bigger, so there is the interesting possibility
that gravitational scattering and black hole production will lead to a
sizeable event rate, higher than in the SM. 

The shape of the shower is probably one of the better ways to
characterize these processes. In the SM charged-current process, a
significant fraction of the neutrino energy is released to just one or a
few hadrons from the brakdown of the target proton. The shower then
builds up from the cascading hadronic interactions of these few hadrons.
In the scenarios we are considering, the production of a black hole of
mass $M_{bh}\sim \sqrt s=\sqrt{2M_\nu m_p}$ is followed by its very
quick evaporation by emission on the brane \cite{EHM3} of a number of
particles of the order of $M_{bh}/T_{bh}\sim (\sqrt
s/M_D)^{(n+2)/(n+1)}$. For the energies we are considering this number
can be bigger than 100. Then the shower builds more quickly than for SM
processes. It is reasonable to expect that the shapes will differ, very
much in the way that a shower formed by a primary iron nucleus differs
from the shower formed by a primary proton. To investigate the
difference in the case at hand requires a more detailed
study.\footnote{See \cite{GA}.} Note that the BH cross section plotted in
Fig. 3 is inclusive over the mass of the BH. A significant portion of
that cross section is due to the production of not so heavy BH's,
through scattering with partons with small $x$. Moreover, as discussed
above, the cross section $\sigma_{bh}$ is of the same order as the
elastic gravitational scattering at $y\sim 1$. In the latter processes a
significant fraction of the neutrino energy is transferred to a few
proton fragments. We then expect the resulting shower to resemble those
induced by SM physics. In order to assess how well can one distinguish
gravity induced showers from SM showers requires to take into account
all these facts. This is an important point: if an excess of horizontal
shower is observed, the shower shape information will be crucial to
secure that the excess is not due to an underestimate of the (unknown)
neutrino flux.

Finally, in order to establish which process (gravitational elastic
scattering, or black hole production) dominates the signal, one needs
some knowledge about the energy dependence of the incoming neutrino
flux. We have already established that BH production dominates energy
loss. However, as the eikonalized cross section grows with $E$, if the
neutrino flux $J(E)$  decreases with $E$ slowly enough, the number of
BH events at energy $E$ may be overshadowed by soft
scattering events due to neutrini with energy $\gg E$. The signal is
the number $dN(E)$ of showers with energy between $E$ and $E+dE$. In
terms of the neutrino flux $J(E)$ and the differential 
cross section we can write
\begin{equation}
\frac{dN(E)}{dE}\propto \int_E^{E_{max}} 
\frac{dE'}{E'}\, J(E')\,\frac{d\sigma}{dy}(E',y\equiv
\frac{E}{E'}).
\end{equation} $E_{max}$ represents the energy at which $\sigma_{bh}$
becomes larger than the inverse horizontal depth ${x_h}^{-1}$. Neutrini with
$E>E_{max}$ interact right away and cannot generate horizontal showers.
We have studied the above integrand by assuming $J(E)\sim E^\alpha$. 
We found that, in the cases of interest, already for $\alpha< -2$ the signal is  
dominated by
events with large $y$, and then by black holes. 
(More precisely we find that the critical $\alpha$'s for $n=2$, 3 and 6
are respectively equal to $-1.76$, $-1.65$ and $-1.48$.)
 This condition
is satisfied for the cosmogenic neutrino flux in fig. 1 of
ref. \cite{Tyler:2001gt} for which $\alpha\simeq -3$. If the cosmogenic
neutrini dominate the flux, then black hole production and gravitational
scattering at $y \sim 1$, and not the softer processes, will dominate the signal in
horizontal air showers.

To conclude, we hope to have convincingly established that neither
higher-dimensional graviton-mediated neutrino-nucleon scattering nor
black hole production in TeV-gravity models can explain the observed
cosmic ray showers above the GZK limit. Nevertheless, horizontal air
showers may probe these scenarios. In this case, black hole production
and gravitational deflection by a large angle
will be the processes that dominate the signal.

\section*{Acknowledgements}

We would like to thank Gian Giudice, Michael Kachelriess, Hallsie Reno and Alessandro
Strumia for valuable conversations.
RE acknowledges partial support from UPV grant 063.310-EB187/98 and
CICYT AEN99-0315. MM acknowledges support from MCYT FPA2000-1558 and
Junta de Andaluc\'\i a FQM-101.

\end{document}